\documentclass[a4paper,fleqn,usenatbib]{mnras}

\usepackage{newtxtext,newtxmath}
\usepackage[T1]{fontenc}
\usepackage{ae,aecompl}

\usepackage{graphicx}	% Including figure files
\usepackage{amsmath}	% Advanced maths commands
\newcommand{\fe}{{$^{60}$Fe~}}

\newcommand{\HI}{ H{\sc{i}}~}
\newcommand{\NaI}{ Na{\sc{i}}~}

\newcommand{\nbr}[1]{\left( #1 \right)} % normal bracket
\renewcommand{\deg}{^\circ}
\newcommand{\nuc}{\nu_\mathrm{c}}
\newcommand{\GeV}{\mathrm{GeV}}
\newcommand{\ff}[2]{\nbr{\frac{#1}{#2}}}
\newcommand{\cvec}[3]{\nbr{\begin{array}{c} #1\\#2\\#3\end{array}}}

\newcommand{\rva}[1]{#1}
\newcommand{\rvb}[1]{#1}
\newcommand{\fdir}{./}

\newcommand{\fext}{.png}

\title[]{Can the Local Bubble explain the radio background?}
\author[M. G. H. Krause et al.]{
Martin G. H. Krause\thanks{E-mail: M.G.H.Krause@herts.ac.uk} and
Martin J. Hardcastle \\
Centre for Astrophysics Research, School of Physics, Astronomy and Mathematics, University of Hertfordshire, 
College Lane, \\ \:\,Hatfield, Hertfordshire AL10 9AB, UK\\
}

\date{Accepted XXX. Received YYY; in original form ZZZ}

\pubyear{2019}

\begin{document}
\label{firstpage}
\pagerange{\pageref{firstpage}--\pageref{lastpage}}
\maketitle

% Abstract of the paper
\begin{abstract}
%Background emission on approximately the same level independent of the
%direction on the sky has been found throughout the electromagnetic
%spectrum. This has led to the identification of interesting
%populations of astronomical objects as well as the verification of big
%bang theory.
The ARCADE~2 balloon bolometer along with a number of other instruments
have detected what appears to be 
a radio synchrotron background at frequencies below about 3~GHz. 
Neither extragalactic radio
sources nor diffuse Galactic emission can currently account for this finding. 
We use the locally measured Cosmic ray electron population, 
demodulated for effects
of the Solar wind, and other observational constraints combined 
with a turbulent magnetic field model to predict
the radio synchrotron emission for the Local Bubble. 
We find that the spectral index of the modelled radio emission
is roughly consistent with the radio background.
Our model can approximately reproduce the observed antenna temperatures 
for a mean magnetic field strength \rva{$B$} 
between 3-5~nT. We argue that this would not violate observational constraints from
pulsar measurements. However, the curvature in the predicted spectrum
would mean that other, so far unknown sources would have to contribute below 100~MHz.
Also, the \rva {magnetic} energy density would then
dominate over thermal and cosmic ray electron energy density, likely 
causing an inverse magnetic cascade with large variations
of the radio emission in different sky directions as well as high
polarisation. 
We argue that this disagrees with several observations
and thus \rva{that} the magnetic \rva{field is probably much lower, quite possibly
limited by equipartition with
the energy density in relativistic or thermal particles ($B=0.2-0.6$~nT). In the latter case,} 
 we predict a contribution 
\rva{of the Local Bubble to} the unexplained radio background \rva{at most at the per cent level}. 
\end{abstract} % 227 words counted on 8 July 2020

% Select between one and six entries from the list of approved keywords.
% Don't make up new ones.
\begin{keywords}
radio continuum: general --
radio continuum: ISM --
ISM: bubbles --
Galaxy: local interstellar matter --
cosmology: diffuse radiation
\end{keywords}

%%%%%%%%%%%%%%%%%%%%%%%%%%%%%%%%%%%%%%%%%%%%%%%%%%
%%%%%%%%%%%%%%%%% INTRODUCTION %%%%%%%%%%%%%%%%%%
%%%%%%%%%%%%%%%%%%%%%%%%%%%%%%%%%%%%%%%%%%%%%%%%%%
\section{Introduction}\label{sec:intro}
The balloon-borne precision bolometer ARCADE 2 has reported an excess
emission above the Cosmic microwave background (CMB) of $54\pm6$ mK at 3~GHz
\citep{Fixsea11}. 
Together with measurements from the Long Wavelength Array at 40-80
MHz and other measurements \citep{DT18}, this forms the extragalactic radio
background, which dominates the sky emission below 1~GHz. 
When the contributions from the CMB and the Milky Way are removed,
an isotropic component with a power law spectrum
 with index -2.58 when plotting antenna temperature vs. frequency remains
 ($\alpha =0.58$ for flux density $S\propto \nu^{-\alpha}$). The relevant frequency
 range includes the 60-80~MHz region, where the 21~cm signal from the epoch
 of reionisation is expected. An absorption feature of less than 1 per cent of the 
 radio background emission has indeed been found 
 by the Experiment to Detect the Global Epoch of Reionization Signature (EDGES)
  at these frequencies \citep{Bowmea18}. For the interpretation of the absorption feature
  as of cosmological origin, it is important 
  to understand whether the radio synchrotron background is produced locally
  or at high redshift \citep[e.g.,][]{Monsalvea19,ECL20}.

Since the contribution from the Milky Way has a distinct geometry and
is accounted for already in the aforementioned results, the most
straightforward explanation would be a large population of known
extragalactic radio sources, namely radio loud active galactic nuclei
and star-forming galaxies.  At 3~GHz, measurements with the 
Karl G. Jansky Very Large Array find a
combined antenna temperature for all such sources of 13~mK,
significantly below the ARCADE 2 result \citep{Condea12}.
 A similar measurement has
recently been performed with the Low-Frequency Array (LOFAR) with the similar result
that only about 25 per cent of the radio background can be accounted
for by resolved radio sources 
\citep{Hardcastlea21aph}.
%(Hardcastle et al., submitted). 
Another suggestion that has been put forward is a Galactic halo
of cosmic ray electrons with a scale length of 10~kpc \citep{OS13,SubraCowsik13}.
The required particle population would however also produce X-rays
via inverse Compton scattering, which would violate observational constraints
\citep{Singea10}. Also, such a prominent radio halo would be atypical for
galaxies like the Milky Way \citep{Singea15,Steinea20},
\rvb{even though halos of up to a few kpc at 150~MHz have been found recently \citep{Steinea19}.}
These difficulties have inspired a
number of interesting explanations, including for example free-free emission
related to galaxy formation at high redshift \citep{Liuea19} 
and dark matter annihilation \citep{Hoopea12}. See
\citet{Singea18} for a recent review.

We investigate here a comparatively simple explanation: synchrotron
emission from the Local Bubble. The Local Bubble is a low-density
cavity in the interstellar medium around the Solar system
\citep[e.g.,][]{CR87}
The superbubble was likely formed 
by winds and explosions of massive stars \citep{Breitschwea16,Schulrea18a}.
Hot gas in the bubble contributes significantly to the soft X-ray
background \citep{Snowdea97,Snowdea98,Galea14,Snowden15}.
The boundary is delineated by a dusty shell that has
been mapped with absorption data against stars with known distances
\citep{Lallea14,Snowdea15b,Pelgrea20}. Direct observation of
the likely present neutral hydrogen supershell is difficult against the background
of the Milky Way, but the distinct structure of erosion of the
interface towards a neighbouring superbubble has been observed
\citep{Krausea18b}. Similar features are also known from \NaI and \HI 
absorption studies \citep{Lallea14}. Interaction of cosmic ray
particles with the supershell may explain the
high-energy neutrinos observed with IceCube \citep{AKS18,BKS20}.
The superbubble contains high ionisation species \citep{BdA06},
filaments and clouds
of partially neutral and possibly even molecular gas 
\citep[e.g.,][]{GryJen14,GryJen17,RedLin08,RedLin15,Snowdea15a,Farhea19,LRT19}
and is threaded by magnetic fields 
\citep[e.g.,][]{AP06,McComea11,Frischea15,Alvea18,Piirolea20}.
\rva{It has already been suggested as the physical origin of high
latitude radio emission by \citet{Sunea08}.}

We first make an empirical model based on a comparison to the 
non-thermal superbubble in the dwarf galaxy IC~10 (Sect.~\ref{sec:IC10})
and then present a detailed model based on the locally observed 
population of cosmic ray electrons and available constraints
on the magnetic field in the Local Bubble (Sect.~\ref{sec:dmodel}).
We discuss our findings in the context of the observational constraints
in Sect.~\ref{sec:disc} and conclude in Sect.~\ref{sec:conc} that a dominant
contribution of the Local Bubble to the radio background seems unlikely.

%%%%%%%%%%%%%%%%%%%%%%%%%%%%%%%%%%%%%%%%%%%%%%%%%%%%%%%%%%
\section{Empirical model by comparing to the non-thermal superbubble in IC~10}
\label{sec:IC10}
%%%%%%%%%%%%%%%%%%%%%%%%%%%%%%%%%%%%%%%%%%%%%%%%%%%%%%%%%%
Superbubbles are not usually known to emit a non-thermal radio synchrotron
spectrum. One such object has, however, been identified in the dwarf
galaxy IC~10 \citep{Heesea15}. The reason why it stands out against 
thermal and non-thermal radio emission
of the host galaxy might be an unusually strong explosion, a
hypernova, about 1~Myr before the time of observation. 
Its size is, similar to the Local Bubble, $\sim 200$~pc. 
The radio spectrum is a power law with
the same spectral index as the radio background, $S(\nu)\propto
\nu^{-0.6}$. The observed non-thermal emission is 40~mJy at
1.5~GHz. 

We use these properties of the nonthermal superbubble in IC~10 to estimate
those of the Local Bubble as follows.
First, we scale this by a factor of $f_\mathrm{s}=0.1$ to account for the fact
that likely none of the supernovae that shaped the Local Bubble was a
hypernova. With the given spectral index, this yields a flux density of 
2.7~mJy at 3~GHz. With a distance of 0.7~Mpc to IC~10, we then get a spectral luminosity 
of $1.6\times 10^{17}$ W~Hz$^{-1}$. Assuming a bubble radius
of  $100 f_\mathrm{r10}$~pc, we obtain a volume emissivity of
\begin{equation}
l_\nu = 1.3 \times10^{-39} \ff{f_\mathrm{s}}{0.1} f_\mathrm{r10}^{-3}\,
\mathrm{W\,Hz^{-1}\,m^{-3}}
\end{equation}
Placing the Sun at the centre of such a non-thermal
bubble yields a flux contribution from each shell at distance $r$ of
\begin{equation}
\mathrm{d}S_\nu = \frac{4 \pi r^2 \,\mathrm{d}r \, l_\nu}{4 \pi r^2} = l_\nu \,\mathrm{d}r\, .
\end{equation}
The integral is straightforward and results, for a radius of the Local Bubble
of $100 f_\mathrm{rLB}$~pc in:
\begin{equation}
S_\nu = 4\times 10^5 \ff{f_\mathrm{s}}{0.1} \,f_\mathrm{r10}^{-3}\,  f_\mathrm{rLB} 
\ff{\nu}{3\,\mathrm{GHz}}^{-0.6}\,\mathrm{Jy} \, .
\end{equation}
The antenna temperature follows from this via
$T_\nu = S_\nu c^2 / (8\pi k_\mathrm{B} \nu^2)$, and so
\begin{equation}
T_\nu = 113 \ff{f_\mathrm{s}}{0.1} \,f_\mathrm{r10}^{-3}\,  f_\mathrm{rLB} 
\ff{\nu}{3\,\mathrm{GHz}}^{-2.6}\,\mathrm{mK} \, .
\end{equation}

This overpredicts the radio synchrotron background by a factor of two and 
thus demonstrates that the
contribution of the Local Bubble can in principle be very important.

%%%%%%%%%%%%%%%%%%%%%%%%%%%%%%%%%%%%%%%%%%%%%%%%%%%%%%%%%%
\section{Detailed model of the radio synchrotron emission of the Local Bubble}
\label{sec:dmodel}
%%%%%%%%%%%%%%%%%%%%%%%%%%%%%%%%%%%%%%%%%%%%%%%%%%%%%%%%%%

Thanks to a number of measurements unique to the Local Bubble, it is
possible to predict its radio emission with far better accuracy
than we have done in the previous section. Both
elements required to predict synchrotron emission, the energy
distribution of cosmic ray electrons and positrons and the strength
and geometry of the magnetic field are constrained by recent
experimental data. The Alpha Magnetic Spectrometer (AMS) onboard the
International Space Station (ISS) has measured the near-earth energy
distribution for cosmic ray electrons with energies E between 0.5~GeV
and 1.4~TeV\citep{Aguilea19}. Constraints at lower energy and outside
the volume influenced by the Solar wind have been provided by Voyager~I
\citep{Cummea16}.
The part of this distribution relevant for the radio
background can be calculated once the magnetic field is known, and  
constraints are available from pulsar observations.
We review the observational constraints on both, magnetic field and particle
energy spectrum, in the following three subsections.

%%%%%%%%%%%%%%%%%%%%%%%%%%%%%%%%%%%%%%%%%%%%%%%%%%%%
\subsection{Magnetic field constraints} \label{ss:magfconst}
%%%%%%%%%%%%%%%%%%%%%%%%%%%%%%%%%%%%%%%%%%%%%%%%%%%%
The magnetic field in the local bubble is constrained by measurements of
the Faraday effect, i.e. the rotation of the plane of polarisation of
pulses from radio pulsars, combined with the pulse dispersion as a
function of frequency. Such measurements yield magnetic field
strength estimates of $B = 0.05-0.2$ ~nT \citep{XH19},
but the measurements do not contain information whether this field
strength is volume filling or restricted to a small fraction of the path 
through the Local Bubble.
Field reversals and density inhomogeneities affect the estimate. 
The quantities directly measured from the pulsar measurements are
dispersion measure DM and rotation measure RM. 
For eight pulsars at distances between 90-140~pc, 
i.e., towards the edge of the Local Bubble, \citet{XH19} 
report a mean dispersion measure of 
42~cm$^{-3}$~pc with a standard deviation of 20~cm$^{-3}$~pc.
This corresponds to a column of free, thermal electrons of 
\begin{equation}
N_e=(1.3\pm0.6)\times 10^{24}$~m$^{-2}\,.
\end{equation}
X-ray measurements of the hot bubble plasma suggest a thermal electron
density of $n_{e,\mathrm{X}} = (4.68 \pm0.47) \times 10^3 $~m$^{-3}$ \citep{Snowdea14}.
This value is very typical for superbubbles, including X-ray bright ones,
as shown in 3D numerical simulations \citep{Krausea13a,Krausea14a}.
The contribution to the free electron column in the Local Bubble from
the X-ray emitting plasma, again for a radius of the Local Bubble
of $100 f_\mathrm{rLB}$~pc is therefore
\begin{equation}
N_{e,\mathrm{X}}=(1.4\pm0.1)\times 10^{22}f_\mathrm{rLB}\,\mathrm{m}^{-2}\,.
\end{equation}
Warm clouds within the Local Bubble have sizes of several parsecs and electron densities
of the order of $n_{e,\mathrm{wc}}=10^5$~m$^{-3}$ \citep[e.g.,][]{GryJen17,LRT19}. 
Assuming a total warm cloud path length of 
$10 f_\mathrm{wcp}$~pc, we obtain an estimate for the corresponding
free electron column of:
\begin{equation}
N_{e,\mathrm{wc}}=3\times 10^{22}f_\mathrm{wcp}\,\mathrm{m}^{-2}\,.
\end{equation}
Hence, neither the hot X-ray plasma nor the warm clouds and filaments 
contribute significantly
to the pulsar dispersion measures. As \citet{XH19} note, the dispersion measure
is probably produced predominantly by the bubble wall, an ionised mixing layer
between the superbubble interior and the cold supershell 
\citep[compare also][]{Krausea14a}.

The root mean square rotation measure against the aforementioned eight pulsars
is 33~rad~m$^{-2}$. For a plasma with electron density $n_e$
and line-of-sight magnetic field $B_{\mathrm{los}}$, 
the rotation measure may be expressed as:
\begin{equation}
RM = 8.1 \,\mathrm{rad\, m}^{-2}\,
\int_\mathrm{Observer}^\mathrm{Source} \ff{n_e}{10^{6}\,\mathrm{m}^{-3}} 
\ff{B_\mathrm{los}}{\mathrm{nT}} \,
\frac{\mathrm{d}l}{\mathrm{pc}}\, ,
\end{equation}
where $\mathrm{d}l$ is the path length element.

For the warm clouds, an estimate for the magnetic field strength
is available from measurements of energetic neutral atoms that are thought
to originate from the solar wind, are scattered by the magnetic field near
the heliospheric boundary and experience charge exchange reactions
\citep{McComea11,McComea20}. For the warm clouds 
surrounding the heliosphere this leads to an estimate of 0.3~nT
\citep{SMcC19}. Pressure balance with the volume filling
X-ray plasma generally suggest $\approx 0.5$~nT for warm clouds in the
Local Bubble \citep{Snowdea14}.

Ignoring field reversals yields an upper limit for the rotation measure
for given electron density, magnetic field $B$ and total path length $l_\mathrm{pc}$.
For the warm clouds we write this as:
\begin{equation}
RM < 4\,\mathrm{rad\, m}^{-2}\,\ff{n_e}{n_{e,\mathrm{wc}}} \ff{B_\mathrm{los}}{0.5\,\mathrm{nT}} f_\mathrm{wcp}\, .\label{eq:rmwc}
\end{equation}
This suggests a perhaps non-negligible, but certainly not dominant contribution by the warm
clouds to the rotation measure.
Scaling to the properties of the X-ray plasma, we write eq.(\ref{eq:rmwc}) as:
\begin{equation}
RM < 38\,\mathrm{rad\, m}^{-2}\,\ff{n_e}{n_{e,\mathrm{X}}} \ff{B_\mathrm{los}}{10\,\mathrm{nT}} f_\mathrm{rLB}\, .
\end{equation}
Consequently, the X-ray emitting plasma in the Local Bubble may be magnetised up to a level of
at least 10~nT without violating the rotation measure constraint. Since we show 
below that very small magnetic fields will not lead to an interesting amount
of radio emission, we consider in the following only magnetic field
strengths between 0.1 and 10~nT.

%%%%%%%%%%%%%%%%%%%%%%%%%%%%%%%%%%%%%%%%%%%%%%%%%%%%
\subsection{Constraints on the particle energy spectrum}\label{ss:e-energies}
%%%%%%%%%%%%%%%%%%%%%%%%%%%%%%%%%%%%%%%%%%%%%%%%%%%%
When averaging over the angle between
the magnetic field direction and the isotropically assumed particle
directions, the characteristic frequency for synchrotron emission
becomes \citep{Longair2011}:
\begin{equation}
\nuc=794\,\mathrm{MHz} \ff{E}{\GeV}^2 \ff{B}{\mathrm{nT}} \,. 
\end{equation}
For magnetic field strengths within the observational limits (Sect.~\ref{ss:magfconst}), cosmic
ray electrons from 50~MeV up to about 6~GeV radiate at frequencies relevant to the
radio background (20~MHz to 3~GHz). Particles at these energies are strongly affected by
the solar modulation, i.e. the energy spectrum changes during the
propagation from interstellar space through the magnetised Solar wind
before reaching the detector near Earth. The Voyager~1 spacecraft has
left the region influenced by the Solar wind in 2012 and has since
then measured electron energy distributions in the range 2.7-79 MeV in
the local interstellar medium \citep{Cummea16}. Cosmic ray propagation models
constrained by Voyager 1 and AMS data \citep{Aguilea19} 
have been developed that infer
the cosmic ray electron density distribution in the local interstellar
medium, outside the Solar wind bubble for energies between 1 MeV and 1
TeV \citep{Vittea19}. The resulting distribution can be approximated by  
$n(E)\propto E^{-p}$, with $p=1.4$ (3.1) below
(above) 1 GeV. 
\rva{\citet{Orlando18} derived a very similar electron
energy distribution and showed that the expected inverse Compton emission 
is consistent with gamma-ray observations. 
Positrons, which are to a large part produced by hadronic interactions \citep{SOJ11},  
contribute at a level of several per cent to the all electron energy spectrum 
in the relevant GeV range, and are included in our model. }

Turbulent mixing is expected to homogenise the electron energy spectrum 
throughout the Local Bubble, even though tangled magnetic fields may prevent
free streaming:
The gyroradius is a function of electron energy $E$ and magnetic field $B$ and is given by
\begin{equation}
r_\mathrm{g} = 3\times10^{-7}\,\mathrm{pc}\, \ff{E}{\mathrm{GeV}} \ff{B}{\mathrm{nT}}^{-1}\,. 
\end{equation}
\rva{The cosmic ray electrons relevant to the radio background
would hence have gyroradii between $10^{-9}$~pc and $10^{-5}$~pc.}
The particles are therefore tied to probably 
tangled magnetic field lines locally. Still, mixing is expected to occur due to gas 
sloshing caused by off-centre supernovae \citep{Krausea14a}. The 
characteristic timescale is the
turnover timescale of the bubble, which can be approximated by the sound crossing time
\citep[e.g.,][]{Krausea13b}.
We argue in Sect.~\ref{ssec:mfgeo} that the Local Bubble 
has evolved probably for several crossing times since the last supernova about 1.5-3.2~Myr ago.
\rva{Therefore, cosmic ray electrons produced by that supernova or any source
that contributed on a similar timescale are now well mixed throughout the superbubble.}
 In the following, we use the
electron and positron energy spectra tabulated in \citet{Vittea19} 
\rva{as representative
for the cosmic ray electron energy spectrum in the Local Bubble.}

\begin{figure*}\centering
	\includegraphics[width=\textwidth]{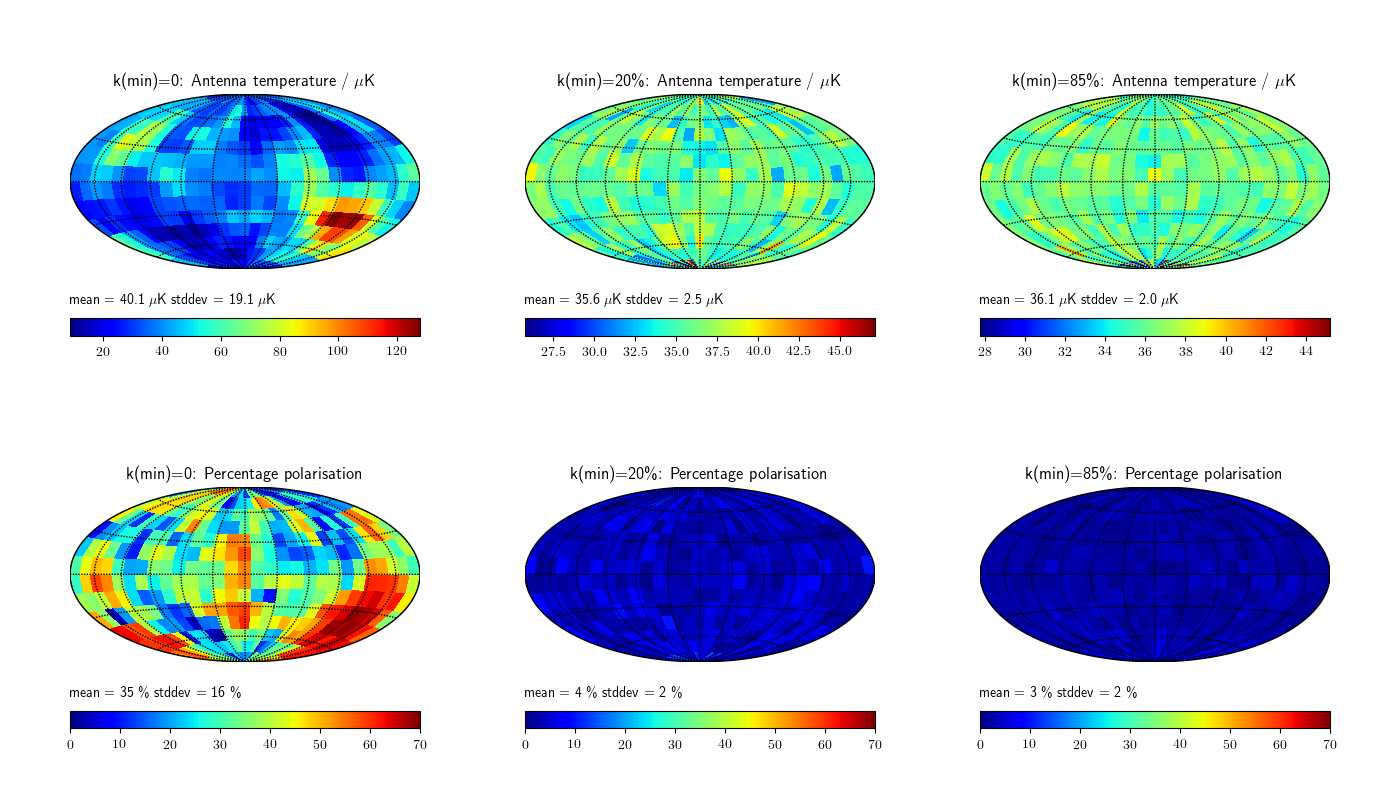}
	\caption{Synthetic radio sky for the detailed Local Bubble model 
	(Sect.~\ref{sec:dmodel}) with a mean magnetic field of 1.6~nT at 3.3~GHz.
	The resolution is 12$^\circ$ matching that of the ARCADE~2 radiometer.
	The top row shows the distribution of the antenna temperature. The bottom row shows 
	the fractional polarisation for the corresponding image. The left column is for a complete
	Kolmogorov power spectrum. The middle (right) one is for a model with the 20 (85) 
	per cent largest modes set to zero.}
    	\label{fig:skysim3GHz}
\end{figure*}

%%%%%%%%%%%%%%%%%%%%%%%%%%%%%%%%%%%%%%%%%%%%%%%%%%%%
\subsection{Constraints on the magnetic field geometry}\label{ssec:mfgeo}
%%%%%%%%%%%%%%%%%%%%%%%%%%%%%%%%%%%%%%%%%%%%%%%%%%%%
The geometry and intermittency of the magnetic field shapes the
directional dependence of the radio synchrotron emission. Supernovae
in superbubbles drive gas sloshing on the scale of the superbubble
diameter, which leads to decaying turbulence \citep{Krausea14a}. Deposits of radioactive
\fe in deep sea sediments suggest that the last supernova in the
Local Bubble occurred 1.5-3.2 Myr ago \citep{Wallnea16}. The characteristic
decay time for turbulence is the sound crossing time. Using a
characteristic diameter of 300~pc \citep{Pelgrea20} and  a sound speed 
of 160~km~s$^{-1}$  
\citep[for an X-ray temperature of 0.1~keV,][]{Snowdea14} gives
a sound crossing time of 1.8~Myr. Superbubbles with sizes comparable to the
Local Bubble may have higher temperatures shortly after 
the supernova explosion \citep{Krausea18b}. 
Therefore, turbulence may have evolved effectively by several decay times 
since the last
explosion. Additional kinetic energy may currently be injected by a
nearby pulsar wind, which is required to explain the observed abundance
of high energy electrons and positrons measured by AMS 
\citep{Lopea18,Bykea19}. 

Observationally, the magnetic field geometry is constrained by
starlight polarisation. For stars with distances 100-500 pc, a
large-scale coherent field is observed towards galactic coordinates
$l=240\deg$--$(360\deg)$--$60\deg$, whereas a magnetic field tangled on small scales is
observed for other longitudes \citep{Berdea14}. The directions with coherent magnetic
field structure appear correlated with the direction towards which the
edge of the Local Bubble is nearest \citep{Pelgrea20}. 
It appears therefore plausible that the coherent structure is a feature of the
bubble wall and that the interior of the Local Bubble has a magnetic field
structure characterised by decaying turbulence, with the largest
magnetic filaments about 40 pc long \citep{Piirolea20}.

%%%%%%%%%%%%%%%%%%%%%%%%%%%%%%%%%%%%%%%%%%%%%%%%%%%%
\subsection{Synchrotron emission model}\label{ssec:em-mod}
%%%%%%%%%%%%%%%%%%%%%%%%%%%%%%%%%%%%%%%%%%%%%%%%%%%%
We therefore model the magnetic field in the Local Bubble as a random
field with a vector potential drawn from a Rayleigh distribution 
with a Kolmogorov power
spectrum following, e.g., \citet{Tribble91} and \citet{Murgea04}. 
We use magnetic field cubes with 256 cells on a side. Most quantities are well 
converged with this resolution. For some we obtain meaningful upper limits 
(compare below).
The approach is well tested for the description of magnetic fields in clusters 
of galaxies with and without radio lobes
\citep[e.g.,][]{Guiea10,HEKA11a,Hardcastle13,HK14}. Following the experimental data on
the field’s geometry, we set the 85 per cent largest modes to zero. 
This is a reasonable approximation for decaying turbulence in the case of
initially weak magnetic fields that were amplified by a strong driving
event \citep{Brandenbea19}, e.g., the sloshing following an 
off-centre supernova explosion \citep{Krausea14a}. The magnetic field 
geometry is discussed further in Sect.~\ref{sec:disc}, below.
We also show models for the uncut power spectrum and for a cut at 20~per
cent for comparison.
We have checked that varying this cutoff has a negligible effect on the
resulting sky temperature \citep[compare][]{Hardcastle13}.

We
put the observer in the centre of the data cube, scale the magnetic
field to values within the range allowed by observations and assume a
homogeneous distribution of synchrotron-emitting leptons. 
%as derived
%from the observations discussed above. 
We derive the density
of non-thermal electrons and positrons, $n_{e,p}$, in the local interstellar medium
at a given energy, from  the tabulated fluxes
$\Phi_{e,p}$ from the model of \citet{Vittea19}. The total density
of non-thermal electrons and positrons, $n(E)$, is then obtained by summing
the individual contributions.

\begin{figure*}\centering
	\includegraphics[width=0.45\textwidth]{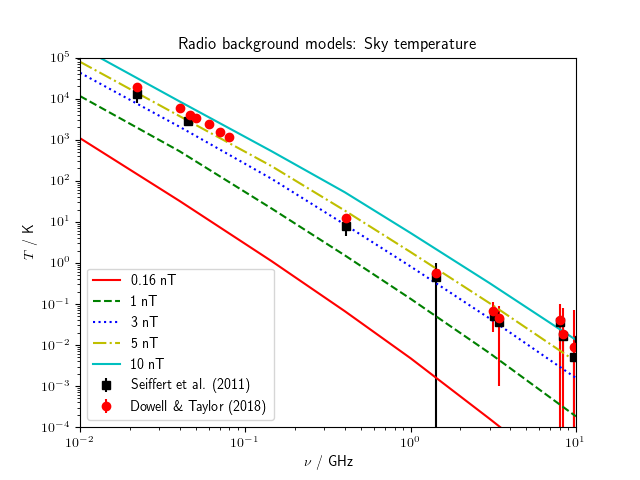}	
	\includegraphics[width=0.45\textwidth]{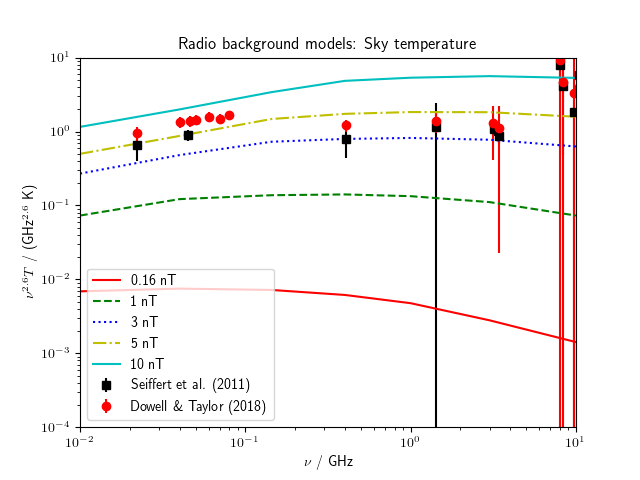}
	\caption{Predicted radio synchrotron emission for the Local Bubble for 
	$k\mathrm{min}=0.85$ (Sect.~\ref{ssec:results}) and different mean
	magnetic field strengths between 0.16~nT (energy equipartition between 
	thermal energy, cosmic ray leptonic internal energy and magnetic energy) and 
	10~nT (conservative limit from Faraday rotation). Measurements are from
	\citet{Seiffea11} and \citet{DT18} as indicated in the legends.
	Left: antenna temperature against 
	observing frequency. Right: Antenna temperature scaled with ($\nu/$GHz)$^{2.6}$).
	A magnetic field strength between 3 and 5~nT is required in the Local Bubble 
	to fully explain the radio background.}
    	\label{fig:LBsynspec}
\end{figure*}

In each energy bin, we use the two neighbouring bins to fit a local power law:
$n(E)=\kappa E^{-q}$. This enables us to use the synchrotron emissivity
for a power law distribution of electrons \citep{Longair2011}:
\begin{equation}
J(\nu) = A\frac{\sqrt{3\pi}e^3B}{16\pi^2\epsilon_0cm_e(q-1)} \kappa\ff{2\pi \nu m_e^3 c^4}{3eB}^{-\frac{q-1}{2}}
\end{equation}
with
\begin{equation}
A=\frac{
\Gamma\nbr{\frac{q}{4}+\frac{19}{12}} 
\Gamma\nbr{\frac{q}{4}-\frac{1}{12}}
\Gamma\nbr{\frac{q}{4}+\frac{5}{4}}
}{\Gamma\nbr{\frac{q}{4}+\frac{7}{12}}}\, .
\end{equation}
Here, $B$ denotes the magnetic field strength perpendicular to the line of sight,
$m_e$ and $e$ are, respectively, electron mass and charge, $c$
is the speed of light and $\epsilon_0$ the vacuum permittivity.
We divide the sky in a $n_\mathrm{lon}\times n_\mathrm{lat}$ grid of longitudes $l$ 
and latitudes $b$ with spacings $\Delta l$ and $\Delta b$. 
For each cone of given $l_i$ and $b_j$, we first select the 
 observing frequency $\nu$.  In each cell, we evaluate the Lorentz
factor given the local magnetic field and the chosen observing frequency. 
We then look up the corresponding
non-thermal electron densities and fit the normalisation and slope
of the local power law at the corresponding energy. After cutting a small region
near the centre of the box (5 per cent of the path length) to avoid resolution effects, 
we find the spectral flux density by summing the weighted emissivities within a given cone:
\begin{equation}
S_\nu (l_i, b_j) = 
\sum_\mathrm{cells\,in\,cone}
\frac{j_\nu \,\mathrm{d}V}{4\pi r}\, ,
\end{equation}
where each Cartesian cell has the same volume d$V$ and $r$ is its distance from the centre 
of the grid, which will be different for each cell.
The intensity is found by dividing through surface area of the corresponding sky grid cell:
\begin{equation}
I_\nu (l_i, b_j) = \frac{S_\nu (l_i, b_j)}{\mathrm{d}l\,\mathrm{d}b\,\sin{b}}\, .
\end{equation}
And, finally, we get the antenna temperature from: 
%Planck's law:
\begin{equation}
%T_{i,j}=\frac{h\nu}{k_\mathrm{B}\ln\nbr{\frac{2 h \nu^3}{c^2 I_\nu (l_i, b_j)} +1}}\, ,
T_{i,j} = \frac{I_\nu (l_i, b_j) c^2}{2 k_\mathrm{B}\nu^2}\, .
\end{equation}
%where, $k_\mathrm{B}$ and $h$ denote, respectively, Boltzmann's and Planck's constant.

We also calculate polarisation information. The local contributions to the Stokes 
parameters are \citep[compare][]{HK14}:
\begin{equation}
\cvec{j_I}{j_Q/\mu}{j_U/\mu} \propto (B_\phi^2 + B_\theta^2)^\frac{q+1}{4} 
\cvec{B_\phi^2 + B_\theta^2}{B_\phi^2 - B_\theta^2}{2 B_\phi B_\theta}\, ,
\end{equation}
where $B_\phi$ and $B_\theta$ are the components of the magnetic
field in spherical coordinates that are perpendicular to the line of sight
at a given location. The maximum polarisation $\mu$ is given by
\begin{equation}
\mu = \frac{\alpha+1}{\alpha+5/3}
\end{equation}
with the spectral index of the radio emission $\alpha = (q-1)/2$.
As $q$ is fitted to for each energy bin, $\alpha$ depends on the
observing frequency. 
The Stokes parameters are integrated along the line of sight to 
obtain $I$, $Q$ and $U$ for each direction of the sky grid.
The fractional polarisation $f$ is then computed as:
\begin{equation}
f = \frac{\sqrt{Q^2+U^2}}{I}\, .
\end{equation}

%%%%%%%%%%%%%%%%%%%%%%%%%%%%%%%%%%%%%%%%%%%%%%%%%%%%
\subsection{Modelling results}\label{ssec:results}
%%%%%%%%%%%%%%%%%%%%%%%%%%%%%%%%%%%%%%%%%%%%%%%%%%%%
The sky distribution of antenna temperature is shown for parameters suitable for 
comparison to the ARCADE~2 experiment in the top row of Fig.~\ref{fig:skysim3GHz}.
The polarisation map for the corresponding model is shown in the bottom row
of the same figure. The observing frequency is 3.3~GHz and the spatial resolution
is 12$^\circ$.

We have chosen three different cuts $k_\mathrm{min}$ in the power spectrum
for the magnetic field (compare Sect.~\ref{ssec:em-mod}). 
The left column is for an uncut Kolmogorov power spectrum. The middle (right)
one for the case where the 20 (85) per cent largest modes are cut.
Large modes in the magnetic power spectrum lead to differences in antenna
temperature of a factor of a few for different sky directions. Consequently,
the standard deviation of the antenna temperature is almost half of the
mean value.
There is little difference between the 
sky distributions predicted for $k_\mathrm{min}=20$~per cent and  
$k_\mathrm{min}=85$ per cent. In both cases, the distribution is smooth across
the sky with maximum antenna temperature ratios below two for any two
sky directions and a standard deviation of less than 10 per cent of the mean.

A noteworthy polarisation signal is only predicted for the full Kolmogorov 
power spectrum. The more the large modes are cut, the lower the polarisation,
again with little difference between $k_\mathrm{min}=20$~per cent and  
$k_\mathrm{min}=85$ per cent, namely 4 per cent  versus 3 per cent.
We note that the polarisation we give for the 
$k_\mathrm{min}=85$ per cent case is an upper limit as this value
was not numerically converged with our largest grid of 256$^3$ cells.

We plot the mean antenna temperature against observing frequency 
in Fig.~\ref{fig:LBsynspec} (left). The Local Bubble has a power law radio spectrum
very similar to that of the radio background (spectral index $\alpha \approx 0.6$). 
We compare to the measurements discussed above and reported by
\citet{Seiffea11} and \citet{DT18}. 
\rva{\citet{Seiffea11} use ARCADE~2 balloon flight and lower frequency radio surveys.
They subtract the Galaxy model from \citet{Kogea11} and an estimated
contribution from external galaxies from the data, and then fit a combination of the 
cosmic microwave background and the radio synchrotron background 
to the remaining spectrum. \citet{DT18} additionally use data from the 
Long Wavelength Array and follow similar methods to obtain the spectrum of the
radio background.}

Good agreement with the data is found 
for magnetic field strengths between 3 and about 5 nT. For more detailed 
comparison to the observations, we remove the $\nu^{-2.6}$ scaling 
in Fig.~\ref{fig:LBsynspec} (right). There is a slight systematic offset
between the two observational data sets, which \citet{DT18}
ascribe to difficulties in the zero-level calibration of low frequency surveys.
There could also be differences due to the removal of the emission of the Galaxy.
This aside, the Local Bubble model also has difficulties in simultaneously fitting
the data points below and above 100~MHz. For example, for the data set
by \citet{Seiffea11}, the 45~MHz data point lies on our 5~nT curve, whereas the
408~MHz data point is on our 3~nT curve. 

For the reference frequency of 400~MHz, our results are well fit by the 
power law:
\begin{equation}
T = 1.44\,\mathrm{K}\,\,\ff{B}{\mathrm{nT}}^{1.62}
\end{equation}

%%%%%%%%%%%%%%%%%%%%%%%%%%%%%%%%%%%%%%%%%%%%%%%%%%%%
\section{Discussion}\label{sec:disc}
%%%%%%%%%%%%%%%%%%%%%%%%%%%%%%%%%%%%%%%%%%%%%%%%%%%%

We used the available data on relativistic particles, magnetic fields,
and thermal components to model the radio synchrotron emission
of the Local Bubble.
We find that the predicted radio spectra show an approximate scaling
of the antenna temperature with frequency as $T\propto \nu^{-2.6}$.
To produce the sky temperature of the ARCADE~2 excess, we 
require a magnetic field in the Local Bubble of
3-5~nT. This is consistent with the pulsar rotation measures, as 
argued in Sect.~\ref{ss:magfconst}, above.

There are, however, some severe difficulties with this solution.
First, the cosmic ray electron spectrum is curved, and this 
translates to a clearly visible curvature in our predicted radio spectra 
(Fig.~\ref{fig:LBsynspec}), but does not show up in the data.
The Local Bubble would of course not be the only contributor
to the radio background. In fact, \citet{Condea12} and Hardcastle
et al. (2020, submitted) both find a contribution of about 25~per cent
of the emission from discrete extragalactic radio sources.
Still, if most of the remaining high frequency emission were explained
by the Local Bubble, it seems that the low frequency data points would require
yet another contributing source. The magnetic field required to explain 
75~per cent of the radio synchrotron background 
(using the 408~MHz data point from \citet{Seiffea11} as a reference value)
would be 2.5~nT.

\rva{At this magnetic field strength, radiative losses are still negligible: For electrons that radiate at a frequency $\nu_\mathrm{c}$, we can write the 
loss timescale due to synchrotron radiation as \citep{GS69}:
\begin{equation}
t_\mathrm{c,sync} = 7\,\mathrm{Myr}\, \ff{\nu_\mathrm{c}}{\mathrm{GHz}}^{-1/2}
\ff{B}{5\,\mathrm{nT}}^{-3/2}\,.
\end{equation}
The dominant radiation field for inverse Compton scattering is expected
to be star light with a wavelength around 1~$\mu$m, where the energy density is 
approximately $U_\mathrm{rad}=6\times10^{-14}$~J/m$^3$ \citep{Popescea17}. The inverse Compton cooling time may 
then be written as \citep{Fazio67}:
\begin{equation}
t_\mathrm{c,iC} = 0.6\,\mathrm{Gyr}\, \ff{E}{\mathrm{GeV}}^{-1}
\ff{U_\mathrm{rad}}{10^{-13}\,\mathrm{J\,m^{-3}}}^{-1}\,.
\end{equation}
These times are long compared to the time since the last supernova, 1.5-3.2~Myr ago
(compare Sect.~\ref{ssec:mfgeo}), a plausible candidate for accelerating
the GeV electrons \citep[compare][]{Sunea08}.
Hence, even in scenarios, where the Local Bubble explains a high fraction of the radio background,
no significant curvature of the radio spectrum would be expected. Gamma-ray
measurements identify a spectral break at an energy around 1~TeV \citep{Lopea18}.
Identifying this break with the break expected from synchrotron cooling fixes the magnetic
field to a value of approximately 0.2~nT.
}

Different magnetic field values mean that different parts of the
particle spectrum are contributing to the observed emission.
Therefore the curvature in the predicted spectra depends on the
magnetic field strength. For magnetic field strengths around and below
1~nT, the curvature would better correspond to the one of the observed
radio background. At this level of magnetic field strength, the Local Bubble
would contribute about 20~per cent of the radio background between
10~MHz and 10~GHz.

The magnetic field strength for equipartition
between magnetic energy and energy in relativistic leptons 
in our Local Bubble model is
$ B_\mathrm{eq,rel}=0.16~nT $.
For equipartition between magnetic and thermal energy, 
using the pressure of $1.5\times 10^{-13}$~Pa given by \citet{Snowdea14},
it is $ B_\mathrm{eq,th}=0.61~nT $.
A magnetic field strength of 1~nT as  discussed in the 
previous paragraph would therefore mean an energetically dominant
magnetic field. This would create tension with our assumption of 
the magnetic power spectrum, because, if the magnetic energy dominates,
one expects an inverse cascade for the magnetic power
\citep{CHB01,BKT15,RepBan17,Sur19}. The power spectrum would then be 
expected to be dominated
by such large modes at the current time of observation.
Therefore, the distributions in the left column in Fig.~\ref{fig:skysim3GHz}
would approximately apply, i.e., we would predict large differences 
of the background emission in different sky directions and significant 
polarisation. Given that the radio background is found as an isotropic 
component in large sky surveys, this seems in tension with 
observations. A magnetic field ordered on large scales also appears
to be in contradiction with the starlight polarisation measurements
discussed in Sect.~\ref{ssec:mfgeo}, where we argued that the 
largest coherent scale for the magnetic field in the Local Bubble
was 40~pc. \rva{We note that \citet{Singea10} have argued against
large-scale patterns in polarisation for the radio background from
WMAP data.}

For decaying turbulence and an initially weak magnetic field, we expect 
magnetic field amplification up to an equilibrium with the kinetic energy
\citep{Brandenbea19}. This growth phase may last several initial crossing
(turnover) times, up to perhaps ten crossing times, depending on the initial
field strength. It is well known that for turbulence in general, the kinetic energy is
converted to thermal energy, also on a timescale comparable to the crossing time.
The Local Bubble may therefore be in a situation close to equilibrium
between magnetic and thermal energy. For this situation, we would predict
a fairly isotropic contribution of \rva{about 10}~per cent to the radio background. 

Of course, the magnetic field might still be lower, perhaps in equipartition
with the cosmic ray electrons or even lower. 
\rva{For a magnetic field strength of 0.16~nT, which interestingly is associated not only
with equipartition between magnetic energy and relativistic leptons, but would also allow
to interpret the break in the electron energy distribution at 1~TeV as due to synchrotron cooling,
the Local Bubble contributes to the radio background at a level of about 1~per cent.}

\rva{For a magnetic field below equipartition with the thermal energy density,
we expect decaying turbulence, which
would lead to a polarisation of at most a few per cent with no coherent 
large-scale pattern in polarisation
(Fig.~\ref{fig:skysim3GHz}). This is very similar to radio polarisation
in the Galactic plane in general \citep{Kogea07}. }

\rva{Summarising, a contribution of the Local Bubble to the radio background
at the per cent level appears most likely.}

This result is perhaps surprising, given the encouraging scalings from the
non-thermal superbubble in IC~10 (Sect.~\ref{sec:IC10}). There is clearly a difference
in the level of non-thermal energy and magnetic energy between the two superbubbles,
and it would be interesting to understand the reasons for this better.

%%%%%%%%%%%%%%%%%%%%%%%%%%%%%%%%%%%%%%%%%%%%%%%%%%%%
\section{Summary and conclusions}\label{sec:conc}
%%%%%%%%%%%%%%%%%%%%%%%%%%%%%%%%%%%%%%%%%%%%%%%%%%%%
We have modelled the radio synchrotron emission of the Local Bubble,
using observational constraints on the energy distribution of
cosmic ray electrons, magnetic fields, X-ray gas and warm clouds and filaments.
We find that in order to explain the radio synchrotron background 
remaining after subtraction of the Galaxy, the cosmic microwave background
and the contribution of known extragalactic point sources  we require a magnetic field
of 2.5~nT. This would be allowed by constraints from Faraday rotation against 
nearby pulsars. However, in this case, the magnetic field would dominate
energetically, and we would expect an inverse cascade, leading 
to large variations of the background emission in different sky direction,
significant polarisation \rva{with} large coherence lengths for the magnetic field,
\rva{and a synchrotron cooling break in the electron energy spectrum below 1~TeV,}
all of which are difficult to reconcile with observations.
\rva{In order to avoid an inverse turbulent cascade associated with large anisotropies
of the radio emission and significant polarisation, the magnetic energy density
should not exceed the thermal one, and to avoid an unobserved cooling break at
electron energies below 1~TeV, the magnetic field should not exceed $\approx 0.2$~nT. For}
this case, we predict a smooth 
emission with low polarisation and \rva{a} maximum contribution  \rva{to}
the unexplained background \rva{at the per cent level}.
This leaves open the possibility that some of the radio background is
produced at very high redshift, which is an important possibility for the
interpretation of the EDGES absorption signal in the context of the
epoch of reionisation.

%-------------------------------------------------------------------
%The Acknowledgements section is not numbered. Here you can thank helpful
%colleagues, acknowledge funding agencies, telescopes and facilities used etc.
%Try to keep it short.
\section*{Acknowledgements}
\rvb{ We thank the anonymous referee for useful comments that helped to improve the manuscript.
 MJH acknowledges support from the UK Science and Technology Facilities Council (ST/R000905/1).}
%%%%%%%%%%%%%%%%%%%%%%%%%%%%%%%%%%%%%%%%%%%%%%%%%%
\section*{Data availability}
The data and code underlying this article are available in the article and in its online supplementary material.

%%%%%%%%%%%%%%%%%%%% REFERENCES %%%%%%%%%%%%%%%%%%

% The best way to enter references is to use BibTeX:

\bibliographystyle{mnras}
%\bibliography{example} % if your bibtex file is called example.bib\bibliographystyle{aa}
\bibliography{/Users/mghkrause/texinput/references}

%%%%%%%%%%%%%%%%%%%%%%%%%%%%%%%%%%%%%%%%%%%%%%%%%%

%\clearpage

%%%%%%%%%%%%%%%%% APPENDICES %%%%%%%%%%%%%%%%%%%%%

% Don't change these lines
\bsp	% typesetting comment
\label{lastpage}
\end{document}